# A single molecule switch based on two Pd nanocrystals linked by a conjugated dithiol

VED VARUN AGRAWAL, REJI THOMAS, G.U. KULKARNI and C.N.R. RAO
Chemistry & Physics of Materials Unit
Jawaharlal Nehru Centre for Advanced Scientific Research
Jakkur P.O., Bangalore 560 064, INDIA

**Abstract**. Tunneling spectroscopy measurements have been carried out on a single molecule device formed by two Pd nanocrystals (dia, ~ 5 nm) electronically coupled by a conducting molecule, dimercaptodiphenylacetylene. The I-V data, obtained by positioning the tip over a nanocrystal electrode, exhibit negative differential resistance (NDR) on a background M-I-M characteristics. The NDR feature occurs at ~ 0.67 V at 300 K and shifts to a higher bias of 1.93 V at 90 K. When the tip is held in the middle region of the device, a coulomb blockade region is observed (~ ±0.3 V).

## 1. Introduction

Molecules were proposed as active electronic devices as early as 1974, by Aviram and Ratner who put forth the concept of unimolecular rectification [1]. Candidates for molecular wires and switches include phorphyrins, phenylenes and thiophenes as well as their polymeric derivatives with extended π-conjugation [2,3]. Organic electronics using conducting polymers has seen a speedy growth in the last three decades but for molecular electronics to come of age, fabrication and measurement techniques would have to reach the nanoscale. With the advent of atomic imaging tools such as scanning tunneling microscopy (STM) and atomic force microscopy (AFM), along with nanolithography and fabrication techniques, it has become possible to carry out reliable measurements on individual molecules in a circuit. There have been several studies of electrical conduction through conjugated oligomers [4,5], typical examples being phenylene ethynylene molecules [6]. An early study based on dc conduction measurements on an Au cluster array cross linked by 1,4-di(4-isocyanophenylethynyl)-2-ethylbenzene showed enhanced conductivity in the nanocrystal film revealing the conducting nature of the molecule [7]. Using STM, Bumm et al [8,9] reported that 4, 4'-di(phenylene-ethynylene)benzothioacetate exhibits a relatively high conductivity compared to neighboring alkanethiols in a mixed self-assembled monolayer. A variant of this method has been tried out by anchoring Au nanocrystals on bifunctional molecules and performing I-V measurements by either STM or conducting AFM [10,11]. Fan et al [12] used a tuning fork tip method and observed features due to negative differential resistance (NDR) in a series of phenylenes. Chen et al [13,14] using the nanopore method, carried out a systematic study of the NDR effect in molecules containing a nitroamine redox center. In another study on molecules of bisthioterthiophene adsorbed on gold electrodes of a break junction, non-linear I-V curves with step-like features have been observed presumably involving conduction across discrete molecular levels [15]. More recently, Reichert et al [16] performed a study to compare I-V characteristics of symmetric and asymmetric phenylenes using the break junction method.

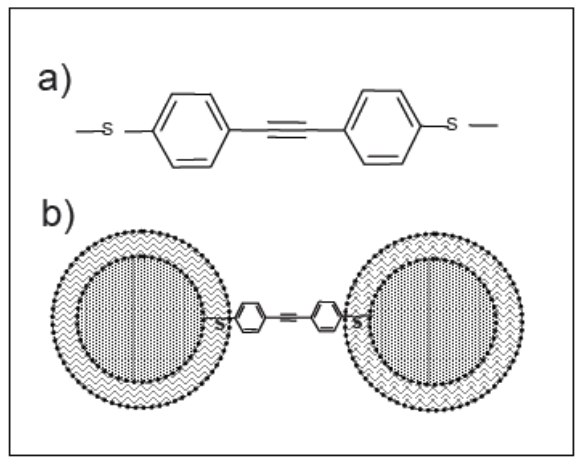

**Scheme 1**: (a) Dimercaptodiphenylacetylene (dmdpa) (b) A schematic showing a dimer of Pd nanocrystals linked by a dmdpa molecule.

In spite of the studies cited above, there are certain unsettled issues related to electrical conduction through a metal-molecule-metal device. While issues such as the nature of molecule-electrode coupling [17,18] and the influence of a non-conducting molecular background in self-assembled monolayer [19] have caused intense debate, ensuring experimentally that only a single molecule actively present between the electrodes remains a problem. As part of continuing interest in electrical transport in molecular nanosystems [20, 21], we have carried out measurements on dimercaptodiphenylacetylene linked to Pd nanocrystals of ~ 5nm diameter on either side (Scheme 1), by employing low temperature STM. We have found a clear signature of NDR which shifts to a higher energy upon lowering the temperature in this single molecule system.

## 2. Experimental

Dimercaptodiphenylacetylene (dmdpa) was prepared by a literature procedure [22] starting with a Sonogashira coupling reaction between 4-iodothioacetate and trimethylsilylacetylene. The resulting derivative was deprotected with tetrabutyl ammonium flouride and coupled with 4-iodothioacetate followed by treatment with NaOH. Nanocrystals of Pd dispersed in ethanol were prepared following the procedure of Miyake et al. [23]. Briefly, a 15 ml of 2.0 mM aqueous solution of $H_2PdCl_4$ was refluxed in a mixture of 10 ml of absolute ethanol and 18 ml of water in the presence of 33.3 mg PVP ($M_w$, 40 000 $gmol^{-1}$). In order to prepare dimeric nanocrystal species linked by the conducting molecule (see Scheme 1), 0.1μl of 2mM solution of dimercaptodiphenylacetylene in toluene was mixed with 1 ml of the Pd sol resulting in a Pd:dmdpa of 5000:1.

Transmission electron microscopy (TEM) was carried out using a JEOL 3010 operating at 300 kV. A drop of the sol treated with the conducting molecule was placed on a copper grid coated with thin carbon film and left to dry overnight. Scanning tunneling microscopy was performed using a SPM 32 (RHK technology, USA) attached to a low temperature stage in an ultra high vacuum chamber. In order to prevent tip induced damage and capture of the nanocrystal by the tip, typically a high impedance of 900 MΩ (bias, 900 mV, set-point current, 1 nA) was used for imaging. The microscope was initially calibrated against the 0001 surface of highly oriented pyrolytic graphite HOPG using an Au tip prepared by electrochemical means [24]. The same tip was used throughout the course of the study. After obtaining a stable non-drifting image, I–V data were collected in the spectroscopy mode by placing the tip atop a feature of interest with the feedback loop turned off. A typical voltage sweep was ± 2 V, with a sweep rate of 10V/s and data interval of 5 mV. Imaging of the area was repeated after the I–V measurement to ensure that the nanocrystals had not drifted away during the I-V measurement. Lower temperatures were achieved by passing liquid nitrogen via a cold stage feedthrough. Temperature was read out using a thermocouple right below the sample mount.

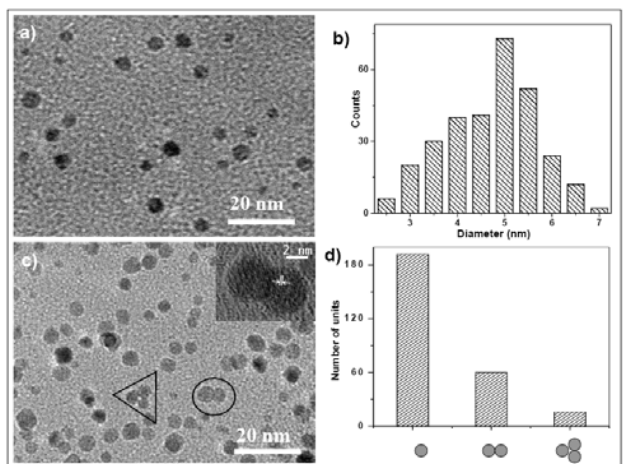

**Figure 1**: a) TEM micrograph of Pd nanocrystals from the as-prepared sol, b) Histogram showing size distribution of the nanocrystal in (a), c) Micrograph taken after adding dmdpa molecules to the Pd sol shows several isolated nanocrystals along with assemblies containing two or three nanocrystals, as marked. Inset shows a high resolution micrograph of a dimeric species. The distance between Pd(111) lattice planes, ~ 2.2 Å. d) Histogram showing the populations of monomers, dimers and trimers.

## 3. Results and Discussion

Figure 1a shows a TEM image revealing isolated near-spherical Pd nanocrystals from the as-prepared sol. The diameters of the nanocrystals is in the range of 3-7 nm as shown in the histogram in Figure 1b, with a mean particle diameter of ~ 5 nm. The TEM image of Pd nanocrystals treated with dmdpa shown in Figure 1c contains some primitive assemblies where two or three nanocrystals. The observed assemblies, dimers and trimers, can only result from the linking of the Pd nanocrystals by the dithiol (dmdpa) molecules. In contrast, the as-prepared sol is completely devoid of such assemblies (Figure 1a). Similar observations have been made previously [25] using bifunctuional molecules. The inset in Figure 1c shows a dimer species with well-resolved {111} planes (d ~ 2.2 Å), although the particle boundaries are not discernible due to line of sight. The populations of the assemblies decrease with the nuclearity (Figure 1d). This is rather an expected trend following a simple estimate based on the known Pd:dmdpa ratio and the assumption that one molecule links a pair of nanocrystals to form a dimer and three for a trimer. It would appear that the probability that a dimer is connected by two dmdpa molecules is negligibly small.

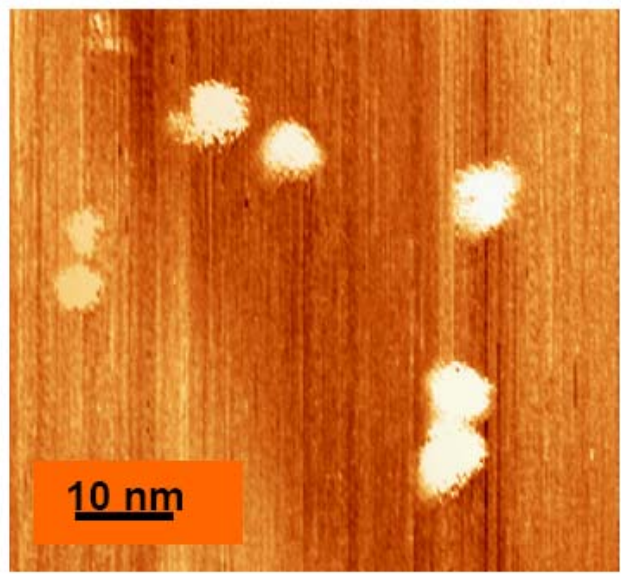

**Figure 2**: STM micrograph of Pd nanocrystals on HOPG substrate.

STM imaging was carried out over large areas (~ 500x500 nm$^2$) initially in order to isolate the monomeric and dimeric species of the nanocrystals. Figure 2 shows a typical STM micrograph of a region containing individual nanocrystals along with a dimer. The nanocrystal diameters are somewhat larger (4 - 8 nm) than the diameters observed in TEM, the difference arising due to the PVP ligand shell around the nanocrystals. Tunneling spectroscopy measurements were carried out on monomers as well as on dimers of nanocrystals. Typical I-V data from an individual nanocrystal at 300 K shows coulomb staircase behavior arising from incremental charging of the nanocrystal with the applied bias (Figure 3). This observation is in accordance with earlier findings on polymer coated metal nanocrystals [20]. I-V measurements on the dimeric species at 300K obtained with the tip atop either of the nanocrystals reveal M-I-M characteristics but for the presence of a sharp feature (see Figure 4a). Around 0.6 V, the current increases rapidly with increasing bias to reach its maximum at ~ 0.67 V and falls sharply thereafter. The presence of such a sharp feature in the I-V data is an indicative of a negative differential resistance (NDR) region in the conductance behavior. The on-off ratio (peak-to-valley ratio) of this molecular device is 2.35. At 90 K, the NDR feature appears at much higher bias of 1.93 V consistent with the temperature dependent behavior of NDR in such molecules [13]. The background current is much less at the low temperature giving rise to a higher on-off ratio of 2.64.

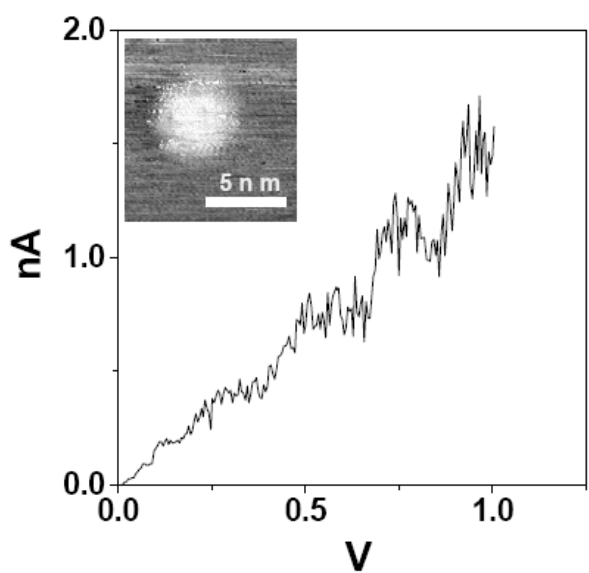

**Figure 3**: I-V data from an isolated Pd nanoparticle showing a coulomb staircase behavior. Inset shows the STM micrograph.

When the tip is brought in the middle region of the nanocrystal dimer possibly right above the conducting molecule, we see an altogether different behavior (Figure 4b). In this case, the I-V data is devoid of an NDR feature but exhibits a distinct coulomb blockade region in the range, -0.3 and 0.25 V. We also observe an overall decrease in the current by at least an order of magnitude compared the case in Figure 4a. These observations can be understood on the basis of a notional circuit consisting of two RC segments representing the PVP covered nanocrystals, interlinked by a molecular resistance (see right side of Figure 4). When the bias is applied on one of the nanocrystals using the STM tip, a voltage drop is expected along the molecular axis which now controls the overlap between the conduction states. An increase in the bias could result in an increased overlap between the conduction states giving rise to a sudden rise in the current till the states flip away at a critical bias (which in this case is, 0.67 V at room temperature) following say, charging or conformational change in the molecule [26]. When the bias is applied to the center of the device (Figure 4b), the two molecular ends carry similar potential and no current is expected to flow in the circuit till the intrinsic gap of the molecule is overcome. This gives rise to the observed coulomb blockade behavior as shown in Figure 4b.

Having demonstrated clearly the switching action of this single molecular device, the merits of the present system deserves some mention. To our knowledge, this is the first study on electrical transport in dimercaptodiphenylacetylene, which is the simplest among the series of phenylene ethynylene oligomers. The molecule is electrically coupled to the Pd metal electrodes (nanocrystals) through Pd-S bonds. It has been suggested that the Pd-S linkage is the best option among many metal-molecule interfaces, The Au-S bond being the worst! [17,18]. The method of making dimeric species of nanocrystals employed by us ensures that the electrical transport measurements indeed pertain to single molecules with a high degree of confidence, thereby avoiding interactions prevalent in a self-assembled monolayer of molecules. Furthermore, we are able to study the electrical behavior along the molecular axis. Thus, a coulomb blockade (~ ± 0.3 V) occurs when the tip bias is applied to the central region of the molecule. The switching action of the device is evident in the form of the NDR feature when the bias is applied to one of the nanocrystal electrodes.

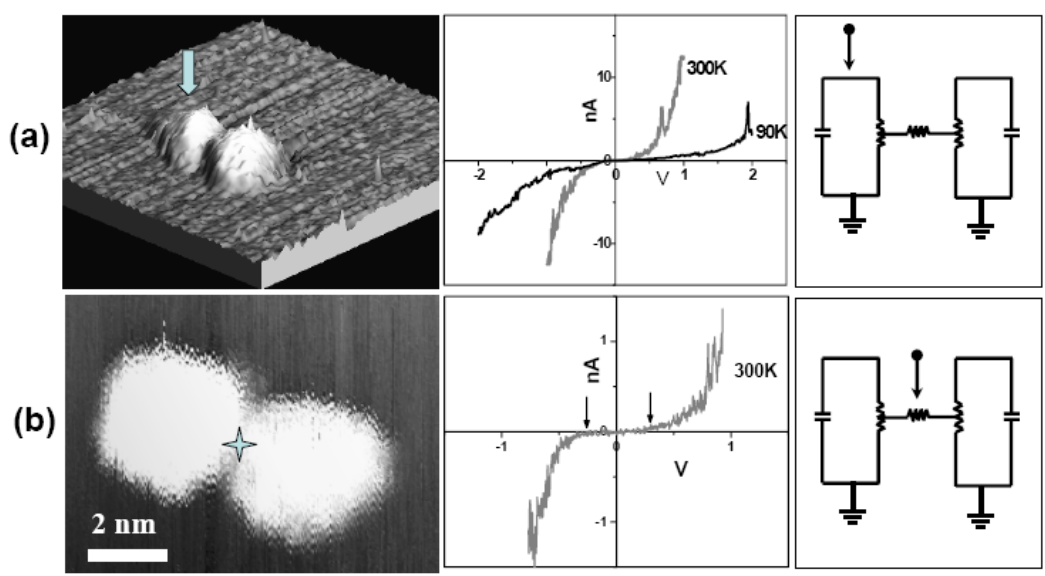

**Figure 4**: Tunneling microscopy and spectroscopy on a dimer of nanocrystals at 300 and 90 K. STM micrographs are shown on the left, I-V data in the middle and schematic circuits on the right side, a) an isometric view, the tip bias being applied to one of the nanocrystals is shown with an arrow, b) a top view with tip bias applied to the central region, as marked.